\documentclass[11pt,a4paper]{article}
\pdfoutput=1
\usepackage{amsmath}
\usepackage{amsfonts}
\usepackage{amssymb}
\usepackage{graphicx}
\usepackage{cite}
\usepackage{subfigure}
\usepackage{comment}
\usepackage{url}
\usepackage{axodraw}
\usepackage{rotating}

\newcommand{\tb}{\tan\beta}
\newcommand{\MS}{M_{S}}

\begin{document}
\title{
\vspace{-1.0cm} 
{\normalsize  DCPT/12/124; IPPP/12/62 \hfill\mbox{}\hfill\mbox{}}\\
\vspace{1.5cm} 
\LARGE{\textbf{Optimising Stop Naturalness
}}}

\author{\vspace{1.5mm}Chris Wymant \\
\small{\em Institute for Particle Physics Phenomenology} \\
\small{\em Department of Physics, Durham University} \\
\small{\em Durham DH1 3LE, United Kingdom}\\
[0.5ex]
}
\date{}
\maketitle

\begin{abstract}
In supersymmetric models a large average stop mass $M_S$ is well known to both boost the lightest Higgs boson mass $m_{h}$ and also make radiative electroweak symmetry breaking unnaturally tuned.
The case of `maximal mixing', where the stop trilinear mixing term $A_t$ is set to give $A_t^2/M_S^2 = 6$, allows the stops to be as light as possible for a given $m_h$.
Here we make the distinction between minimal $M_S$ and optimal naturalness, showing that the latter occurs for less-than-maximal mixing.
Lagrange constrained optimisation reveals that the two coincide closely in the Minimal Supersymmetric Standard Model (MSSM) -- optimally we have $5 < A_t^2/M_S^2 < 6$.
We discuss why the two are not generally expected to coincide beyond the MSSM, and that even within the MSSM different models should not be compared based on the $M_S$ necessary to achieve a given $m_{h}$.
The splitting between the two stop-mass eigenvalues $m_{\tilde{t}_2} - m_{\tilde{t}_1}$ is shown to be unconstrained by naturalness considerations.
\end{abstract}

\section{Introduction}

Radiative electroweak symmetry breaking in the Minimal Supersymmetric Standard Model (MSSM) occurs when the mass matrix for the CP-even Higgs scalars develops a negative eigenvalue.
For moderate or large $\tan\beta$ this is due to the soft supersymmetry (SUSY) breaking mass-squared for the $H_u$ superfield, $m^2_{H_u}$, driven by radiative corrections to be more negative than the supersymmetric Higgs mass-squared $|\mu|^2$.
A nonzero mass for the $Z$ boson results, according to
\begin{equation} \label{REWSB}
-\tfrac{1}{2} M_Z^2 = m^2_{H_u} + |\mu|^2 + \mathcal{O}((\tb)^{-2}).
\end{equation}
From Eq. \eqref{REWSB} we can make statements about naturalness.
A measure commonly used is that of Barbieri and Giudice \cite{Barbieri198863}, which can be calculated for UV-complete models described by a set of fundamental parameters $p_i$.
For a given point in $p_i$ space that results in the observed value of $M_Z$, one calculates derivatives of $\log M_Z$ with respect to $\log p_i$; the observed $M_Z$ is taken to be natural if all such derivatives are $\lesssim \mathcal{O}(1)$.
However this measure does not penalise a situation that we should still regard as unnatural.
First we note that the strong renormalisation-group running of $m_{H_u}^2$ from the high scale $\Lambda$ (where SUSY breaking is mediated to the MSSM) to the electroweak scale, i.e the {\it radiative} nature of the symmetry breaking, motivates a re-writing of Eq. \eqref{REWSB} as 
\begin{equation} \label{REWSB2}
-\tfrac{1}{2} M_Z^2 = m^2_{H_u}(\Lambda) + \delta m^2_{H_u} + |\mu|^2 + \mathcal{O}((\tb)^{-2}).
\end{equation}
It may be the case that at the high scale a single parameter $p$ sets both $m^2_{H_u}(\Lambda)$ and the masses involved in the radiative correction $\delta m^2_{H_u}$, in such a pattern that at the low scale these terms accidentally cancel, even if separately large, regardless of the value of $p$: this is called {\it Focus Point SUSY}.
This parameter then does not need to be finely tuned.
However the cancellation depends sensitively not only on this mass-setting pattern but also on the value of the top Yukawa\footnote{
The squared top Yukawa is a prefactor to the $\delta m^2_{H_u}$ -- see Section \ref{One-loop} -- so cancellation between $m^2_{H_u}(\Lambda)$ and  $\delta m^2_{H_u}$ to $1$ part in $N$ happens only for an {\it ad hoc} tuning of the top quark mass to $1$ part in $2N$.
}, and weakly on the scale $\Lambda$; unless these three are linked by some symmetry the cancellation is accidental.
A natural theory, by contrast, does not have large cancellations except those enforced by symmetries.

A stricter criterion for naturalness is simply to ask that none of the terms contributing to the right-hand side of Eq. \eqref{REWSB2} are dramatically larger than the Z mass squared, \`{a} la Kitano-Nomura \cite{Kitano:2005wc}.
This has the further advantage of allowing bottom-up deductions to be made, i.e. without knowing the underlying high-scale theory, for example as was done in \cite{Papucci:2011wy} to lay out requirements on a natural spectrum.

The stop is the chief contributor to $\delta m^2_{H_u}$, and thus we require light stops for naturalness.
The impressive constraints on the squarks of the first two generations now with $\sim5 \text{fb}^{-1}$ of data (see e.g. \cite{ATLAS-CONF-2012-033,ATLAS-CONF-2012-041}) are considerably relaxed for squarks of the third generation, due to direct production cross-sections suppressed by parton distribution functions and the less distinctive final state signals that may result (e.g. being too similar to Standard Model (SM) top backgrounds \cite{Lee:2012sy}).
Indeed stops decaying to stable neutralinos can still be lighter than the top quark \cite{stopSearches}.
Recent discussions of stop limits and discovery potential can be found in \cite{Cao:2012rz,Essig:2011qg,Kats:2011qh,Bi:2011ha,Desai:2011th,Choudhury:2012kn,Carena:2012np,Han:2012fw,Kaplan:2012gd,Alves:2012ft,Plehn:2012pr,Bai:2012gs}.
The weakening of the bounds for stops and sbottoms is of course only useful for naturalness in the context of models where the third-generation squarks are lighter than those of the first two -- the {\it Effective} or {\it Natural} SUSY paradigm (introduced in references \cite{Dimopoulos:1995mi,Cohen:1996vb} and recently revisited in \cite{Papucci:2011wy,Brust:2011tb}).
The authors of \cite{Papucci:2011wy} argued that this effect ought be a feature of the mediation of SUSY breaking rather than a renormalisation group (RG) running effect, since the same coupling that induces an RG splitting of the squark masses also gives strong running $m^2_{H_u}$ -- precisely what we are trying to avoid\footnote{
An exception would be a heavy Right-Handed sbottom at large $\tan\beta$, which would cause the Left-Handed stop to run lighter than the LH sup and scharm without driving the running $m^2_{H_u}$.
However it is the integral of the running stop masses that gives $\delta m^2_{H_u}$, and so starting heavy but running light only half solves the problem; furthermore we need both stops to be light, not just one.
We thus do not consider this possibility to contradict the argument of \cite{Papucci:2011wy}.
}.
Models that achieve this include \cite{Craig:2012di,Craig:2012hc} where SUSY breaking occurs via gauge mediation, with the Standard Model gauge group supplemented by a progressively broken flavour gauge symmetry \cite{Craig:2012di} or replaced by two separate SU(5) groups in the UV \cite{Craig:2012hc}.

The stop mass and stop mixing also contribute radiatively to the physical mass of the lightest CP-even Higgs boson $h$.
At tree level $m_{h}$ is bounded from above by $M_Z\cos 2\beta$, and saturates this bound in the `decoupling limit' where the pseudoscalar $A$ is appreciably heavier than $M_Z$.
In this limit $h$ also has SM-like couplings, which are favoured by the current signal strength data\footnote{
The diphoton `excess' is downplayed in \cite{Baglio:2012et}, where QCD uncertainties in gluon fusion show the discrepancy is only 1$\sigma$, and also in \cite{Plehn:2012iz}, based simply on an alternative analysis of the coupling data.}
for the $126$ GeV resonance \cite{:2012gk,CMS} assumed to be a Higgs boson.
It has long been known that, in the MSSM at least, some substantial combination of stop mass- and stop mixing-induced corrections to $m_{h}$ is needed to lift it above the LEP lower bound of $114.4~\text{GeV}$.
A $\sim\!\!126~\text{GeV}$ Higgs requires these corrections to be even more substantial, with correspondingly worse implications for naturalness.
The interplay of parameters for such a Higgs mass when looking agnostically at the MSSM has been investigated in \cite{Hall:2011aa,Baer:2011ab,Heinemeyer:2011aa,Arbey:2011ab,Draper:2011aa,Carena:2011aa,Cao:2011sn,Kang:2012tn,Desai:2012qy,Cao:2012fz,Lee:2012sy,Christensen:2012ei,Brummer:2012ns,Badziak:2012rf,CahillRowley:2012rv,Arbey:2012dq,Baer:2012up,Antusch:2012gv}.
Even remaining in the field of supersymmetry many more works have shown the implications of such a Higgs mass in particular models of SUSY breaking, in extensions of the MSSM, or else have focused predominantly on issues relating to the decays of the Higgs into different final states.

In this work we focus on the effects of the stop sector on the physical Higgs mass $m_{h}$ and on the `unnaturalness' term $\delta m_{H_u}^2$; specifically on the relationship between them.
Recently there has been much focus in the literature on the case of {\it maximal mixing}, in which the stop trilinear parameter $A_t$ is related to the average stop mass $\MS$ by $x\equiv A_t^2/\MS^2 = 6$.
One can see that this maximises the stop-mixing contribution to $m_h$, allowing minimal stop masses.
However $A_t$ also contributes to $\delta m_{H_u}^2$; perhaps a smaller value should have been chosen, with heavier stops instead.
Would this have been as good?
We solve the problem analytically with Lagrange constrained minimisation to reveal the optimum balance between $\MS$ and $x$.
A more approximate method (and transparent result) is laid out in Section \ref{One-loop}.
Higher-order complications are considered in Section \ref{Two-loop}.
Section \ref{Conclusion} contains a brief discussion and conclusions.

\section{Leading-Order Analysis} \label{One-loop}

The one-loop beta function of $m_{H_u}^2$ \cite{Martin:1997ns} is:
\begin{equation} \label{mHuBetaFunction}
\begin{split}
16 \pi^2 \frac{d}{dt}m_{H_u}^2 = & \quad 6 y_t^2 (m_{\tilde{Q}_3}^2 + m_{\tilde{u}_3}^2 +A_t^2) \\
& + 6 y_t^2 m_{H_u}^2 -6g_2^2 M_2^2 -\frac{6}{5}g_1^2 M_1^2 + \frac{3}{5}g_1^2 \text{Tr}[Y_{\tilde{f}} m_{\tilde{f}}^2],
\end{split}
\end{equation}
where $t=\log(Q/\Lambda)$, with $\Lambda$ the high/mediation scale at which the soft SUSY-breaking mass terms are generated.
One can roughly neglect the terms of the second line\footnote{
The effect of $m_{H_u}^2$ on its own running is small if the leading log approximation is valid (i.e. $(\mbox{one-loop factor}) \times \log(\Lambda/\MS) < 1)$. Then, since the overall radiative correction must be substantial enough to turn $m_{H_u}^2$ negative, the $m_{H_u}^2$ term in the beta function must be appreciably smaller than the other terms.
The electroweak couplings we neglect, dominated as they are by $y_t^2$.
While the trace term is a sum over all scalars, it couples only through $g_1$ and is `relatively small in most known realistic models' \cite{Martin:1997ns}.
For example it vanishes at the high scale in all models of General Gauge Mediation \cite{Meade:2008wd}, and all models with universal scalar masses (such as minimal supergravity) since $\text{Tr}[Y] = 0$.
Furthermore the running of the trace is proportional to the trace itself.
The wino term on the other hand may be appreciable \cite{CahillRowley:2012rv}, but here we will be differentiating with respect to stop-sector terms, so this effect drops out.
}; keeping only the large stop-sector terms, taking these to be constant and integrating gives the leading log expression
\begin{equation} \label{deltamHuLL}
 \delta m_{H_u}^2 \approx -\frac{3}{8\pi^2}\; y_t^2 \: (m_{\tilde{Q}_3}^2 + m_{\tilde{u}_3}^2 +A_t^2)\log\left(\frac{\Lambda}{\MS}\right),
\end{equation}
at a scale $\MS$ -- the scale at which Eq. \eqref{REWSB} holds most accurately, \cite{Gamberini1990331,PhysRevD.46.3981,deCarlos:1993yy}.

Before connecting Eq. \eqref{deltamHuLL} to the physical Higgs mass $m_h$, we note that it tells us something about stop naturalness on its own.
It can be re-written in terms of the stop mass eigenvalues: taking the stop mass matrix without the subdominant electroweak $D$-term contributions, we have
\begin{equation} \label{deltamHuLL2}
\delta m_{H_u}^2 \approx -\frac{3}{8\pi^2}\; y_t^2 \: \left[m_{\tilde{t}_1}^2 + m_{\tilde{t}_2}^2 - 2m_t^2 + \frac{(m_{\tilde{t}_1}^2 - m_{\tilde{t}_2}^2)^2}{m_t^2} \cos^2\theta_{\tilde{t}}\,\sin^2\theta_{\tilde{t}} \right] \log\left(\frac{\Lambda}{\MS}\right)
\end{equation}
where $\theta_{\tilde{t}}$ is the stop mass mixing angle.
In \cite{Lee:2012sy} it was argued that the final term in square brackets motivates $m_{\tilde{t}_1} \sim m_{\tilde{t}_2}$ for naturalness; then since the left-handed stop shares a mass with the left-handed sbottom ($m_{\tilde{Q}_3}$), non-observation of sbottoms translates into constraints on both stops.
However in Eq. \eqref{deltamHuLL} we can define the average stop mass by $2\MS^2=m_{\tilde{Q}_3}^2 + m_{\tilde{u}_3}^2$, and there is explicit insensitivity to $m_{\tilde{Q}_3}^2 - m_{\tilde{u}_3}^2$ which will split the mass eigenvalues.
The discrepancy arises from the neglected $\cos^2\theta_{\tilde{t}}\sin^2\theta_{\tilde{t}}$ factor in Eq. \eqref{deltamHuLL2}, which goes to zero as we pull apart $m_{\tilde{Q}_3}^2$ and $m_{\tilde{u}_3}^2$.
We see that in fact the two mass eigenvalues can be arbitrarily split without naturalness penalty.

We now ask what Eq. \eqref{deltamHuLL} tells us in conjunction with the physical Higgs mass-squared.
At tree level, in the decoupling limit, the latter is $m_{h,\text{tree}}^2=M_Z^2\cos^2 2\beta$ with dominant radiative correction
\begin{equation} \label{HiggsMass}
 \delta m_{h}^2 \approx \frac{3}{4\pi^2} \frac{m_t^4}{v^2} \left( \log\left(\frac{\MS^2}{M_t^2}\right) + \frac{X_t^2}{\MS^2}\left(1 - \frac{X_t^2}{12\MS^2} \right) \right),
\end{equation}
where $v=174~\text{GeV}$ and $X_t=A_t - \mu \cot\beta$.
The second term is the threshold correction to the Higgs self-coupling $\lambda$ from integrating out both stops at a scale $\MS$; the first term is the SM beta function for $\lambda$ approximately integrated from $\Lambda$ down to a scale equal to the top quark mass, where the Higgs running mass $m_h(t)=\lambda(t)v^2$ coincides closely with the pole mass.

Firstly we note that through Eq. \eqref{REWSB}, the required $|\mu|$ depends on the unknown high-scale value of $m_{H_u}^2$ as well as its radiative corrections.
However, a) the aim for natural SUSY is $|\mu| / (100\mbox{ GeV}) \lesssim \mbox{a few}$, b) a large Higgs mass $\sim\!\!126~\text{GeV}$ needs\footnote{
Unless one enters the realm of split or high-scale SUSY $\MS \gtrsim \mathcal{O}(10^{4,5}\,\mbox{GeV})$, \cite{Giudice:2011cg}.
} $\tan\beta \gtrsim \mathcal{O}(5)$, and c) later we will arrive at $A_t\gtrsim \mathcal{O}(1\mbox{ TeV})$.
Thus we expect $X_t$ to be very close to $A_t$ without knowing the precise value of $\mu$.

Secondly, we see that while the Higgs mass depends only on the average stop mass $\MS$, $\delta m_{H_u}^2$ depends on both $\MS$ and the precise linear combination $m_{\tilde{Q}_3}^2 + m_{\tilde{u}_3}^2$.
We then must choose a definition of $\MS$.
Often this is taken to be a geometric mean; the minimum $(m_{\tilde{Q}_3}^2 + m_{\tilde{u}_3}^2)$ for constant $(m_{\tilde{Q}_3}^2 \times m_{\tilde{u}_3}^2)^{1/2}$ then provides weak motivation for $m_{\tilde{Q}_3}^2 = m_{\tilde{u}_3}^2 $.
If instead the linear average $\MS^2 {\equiv} \tfrac{1}{2}(m_{\tilde{Q}_3}^2 + m_{\tilde{u}_3}^2)$ is chosen, the orthogonal linear combination is entirely free as previously mentioned.
A further alternative would be to take an average of the mass eigenvalues $m_{\tilde{t}_{1,2}}$: the dependence of $\delta m_{H_u}^2$ on the underlying parameters $m_{\tilde{Q}_3}^2,m_{\tilde{u}_3}^2,A_t$ then shifts very slightly but becomes much less transparent, as we have already seen.
We can appeal to the limit\footnote{
$\Lambda/\MS$ is very large in all but the most extreme cases; $m_{\tilde{t}_{2}}/m_{\tilde{t}_{1}}$ cannot be large if we integrate out both stops together to calculate the Higgs mass.
} $\log(\Lambda/\MS) \gg \log(m_{\tilde{t}_{2}}/m_{\tilde{t}_{1}})$, in which the former log and thus $\delta m_{H_u}^2$ has no sensitivity to how $\MS$ is defined.
We can thus take the aforementioned linear average, so that the functions $\delta m_h^2$ and $\delta m_{H_u}^2$ depend on the stop sector simply through $\MS$ and $A_t$.
Note that though other particles besides the stop make smaller contributions to both the Higgs mass and unnaturalness, below we will differentiate with respect to stop-sector parameters and so this effect drops out.

We are now in a position to use Lagrange constrained minimisation: the solution of
\begin{equation} \label{LagrangeMin}
 \frac{\partial}{\partial (\MS^2)} \left(\delta m_{h}^2 - \lambda \, \delta m_{H_u}^2 \right) = \frac{\partial}{\partial (A_t^2)} \left(\delta m_{h}^2 - \lambda \, \delta m_{H_u}^2 \right) = 0,
\end{equation}
where $\lambda$ is the unspecified Lagrange multiplier, is the most natural ratio of $A_t^2$ to $\MS^2$, with the overall scale of one of these two parameters freely chosen thereafter.
Explicitly, with the one-loop $\delta m_h$ \eqref{HiggsMass} and the leading log $\delta m_{H_u}^2$ \eqref{deltamHuLL}:
\begin{equation} \label{MostNaturalRatio1loop}
x_{\text{natural}} \equiv \left(\frac{A_t^2}{\MS^2}\right)_{\text{natural}} = 2 + \sqrt{4+ \frac{6(L-2)}{L-1}} \quad\sim\: 5
\end{equation}
with $L=\log(\Lambda^2 / \MS^2)$.
The solution is real for $L>\tfrac{8}{5}$, asymptotes to $2+\surd 10 \approx 5.16$ as $L\rightarrow \infty$, and is already $5$ for $L=7$ (i.e. $\Lambda / \MS = 33$) -- thus it is essentially constant over phenomenologically interesting mediation scales and stop masses.
That the optimal $x$ should be {\it close} to six is not surprising: using the logarithmic stop mass term to boost the Higgs mass requires exponentially heavy stops and thus exponentially bad fine-tuning; whereas the stop mixing term contribution to $m_h$ can be large even for small $A_t^2$ and $\MS^2$, provided their ratio is favourable.
However the optimal $x$ must in fact be {\it less} than the maximal mixing value $x=6$: decreasing it from $6$ to $6-\delta$ reduces the physical Higgs mass by $\mathcal{O}(\delta^2)$ but increases naturalness by $\mathcal{O}(\delta)$.
We see that {\it almost maximal mixing} is optimal.

\section{Higher-order Effects} \label{Two-loop}

\subsection{Finding $x_{\text{natural}}$}

\begin{figure}[!ht]
\centering
\includegraphics[width=0.55\linewidth]{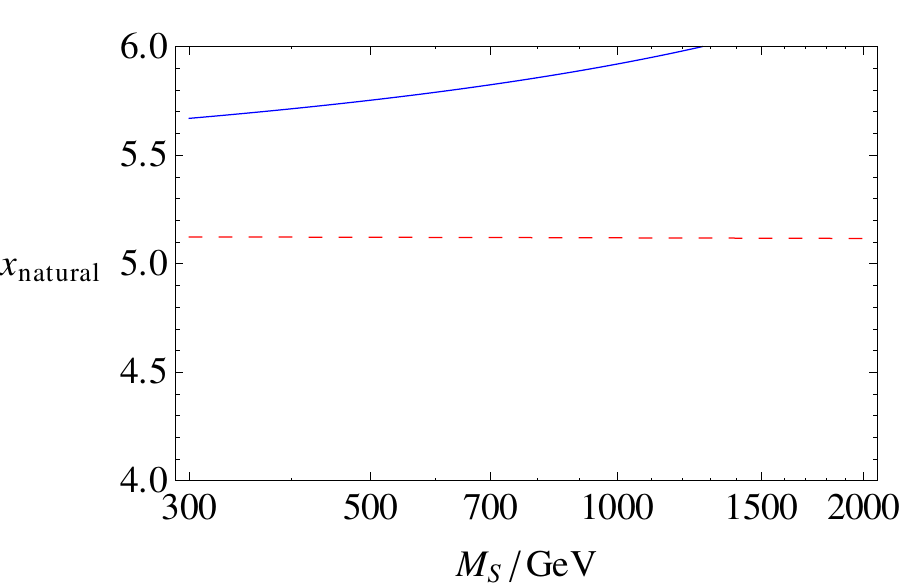}
\caption{The most natural ratio $x\equiv A_t^2/\MS^2$ obtained from maximising the Higgs mass at one loop (red, dashed) and two-loop (blue, solid) for constant electroweak symmetry breaking term $\delta m_{H_u}^2$, as a function of the average stop mass $\MS$.}
\label{fig:RatioNLO}
\end{figure}

Higher order effects of the stop on the physical Higgs mass can be taken into account with the two-loop expression of \cite{Carena:1995bx}:
\begin{gather} 
 \delta m_{h}^2 = \frac{3}{4\pi^2}\frac{m_t^4}{v^2}\left[ \frac{1}{2}\tilde{X}_t + \left( 1+D \right) T +\epsilon\left(\tilde{X}_t T+T^2\right) \right]\,, \label{CarenaWagner} \\
\mbox{with} \quad m_t = \frac{M_t}{1+\frac{4}{3\pi}\alpha_3(M_t)}, \notag \\
\alpha_3(M_t) = \frac{\alpha_3(M_Z)}{1+\frac{23}{12\pi}\alpha_3(M_Z)}, \notag \\
T=\log\frac{ \MS^2}{M_t^2}, \notag \\
D = -\frac{M_Z^2}{2m_t^2}\cos^2 2\beta, \notag \\
\tilde{X}_{t} = \frac{2A_t^2}{\MS^2} \left(1 - \frac{A_t^2}{12 \MS^2} \right), \notag \\
\mbox{and} \quad \epsilon = \frac{1}{16\pi^2}\left(\frac{3}{2}\frac{m_t^2}{v^2}-32\pi\alpha_3(M_t) \right) \notag
\end{gather}
(which also includes the smaller, soft-mass independent, one-loop $D$-term $\mathcal{O}(M_Z^2 m_t^2)$ of \cite{Brignole:1992uf}).
The optimisation, Eq. \eqref{LagrangeMin}, goes through exactly as before.
The solution is the positive root of the following equation (which recovers Eq. \eqref{MostNaturalRatio1loop} as $D,\epsilon \rightarrow 0$)
\begin{multline} \label{MostNaturalRatio2loop}
   \left[  1 + 2\epsilon T + L (-1 + \epsilon -2\epsilon T)\right] x_{\text{natural}}^2 \\
+ 4\left[ -1 - 2\epsilon T + L (1 -3\epsilon +2\epsilon T) \right] x_{\text{natural}} \\
- 6\left[  2 + 4\epsilon T + L (-1 + D -2\epsilon T) \right]= 0
\end{multline}
We show the variation of this solution with $\MS$ in Fig. \ref{fig:RatioNLO}; dependence on $\tan\beta \in [5,45]$ and the top quark mass uncertainty is negligible.

Two other approaches are trivially equivalent to using Eq. \eqref{LagrangeMin} to find $x_{\text{natural}}$.
Firstly, one could invert the $\delta m_{H_u}^2$ expression to find the function $\MS(x)|_{\delta m_{H_u}^2}$ for how the stop mass must vary as a function of $x$ in order to keep $\delta m_{H_u}^2$ constant: from Eq. \eqref{deltamHuLL} this monotonically decreasing function is
\begin{equation}
\MS(x)|_{\delta m_{H_u}^2} = \Lambda\, \exp\!\left(\tfrac{1}{2}\, W_{-1}\!\left( \frac{-16\pi^2\,\delta m_{H_u}^2}{(2+x)\Lambda^2} \right)\,   \right)
\end{equation}
where $W_{-1}(\ldots)$ is the lower branch of the Lambert $W$ function.
The one-parameter function $\delta m_h^2(x,\,\MS(x)|_{\delta m_{H_u}^2})$ then gives the range of Higgs masses possible for a given $\delta m_{H_u}^2$; the {\it maximum} occurs at $x_{\text{natural}}$.

\begin{figure}
\centering
\subfigure{
\includegraphics[width=0.47\linewidth]{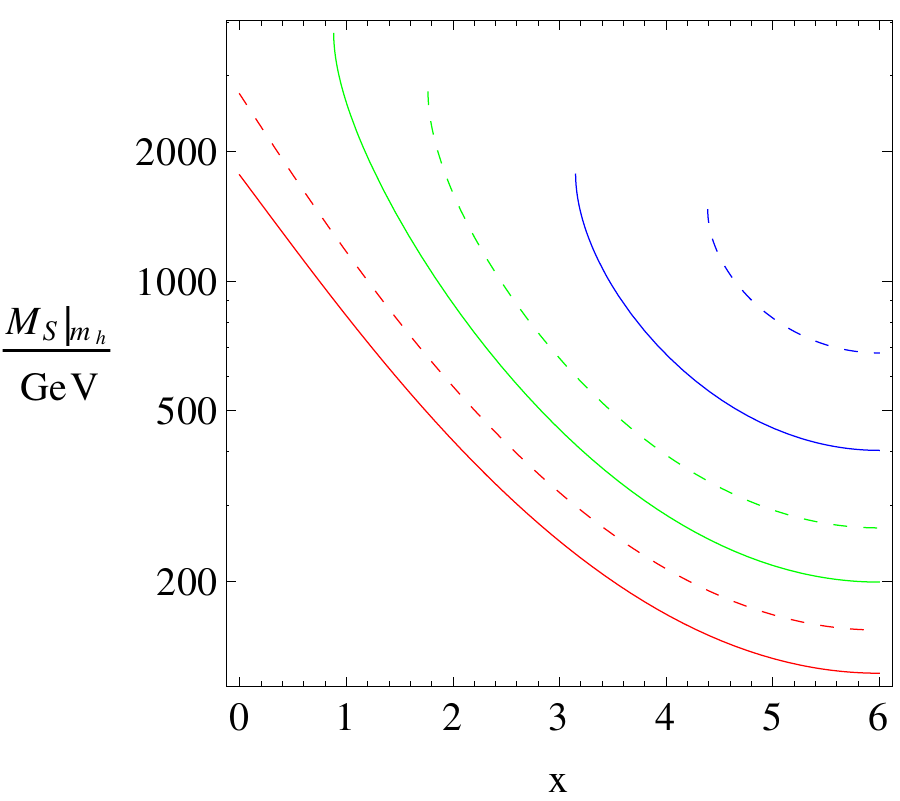}
}
\hspace*{5mm}
\subfigure{
\includegraphics[width=0.43\linewidth]{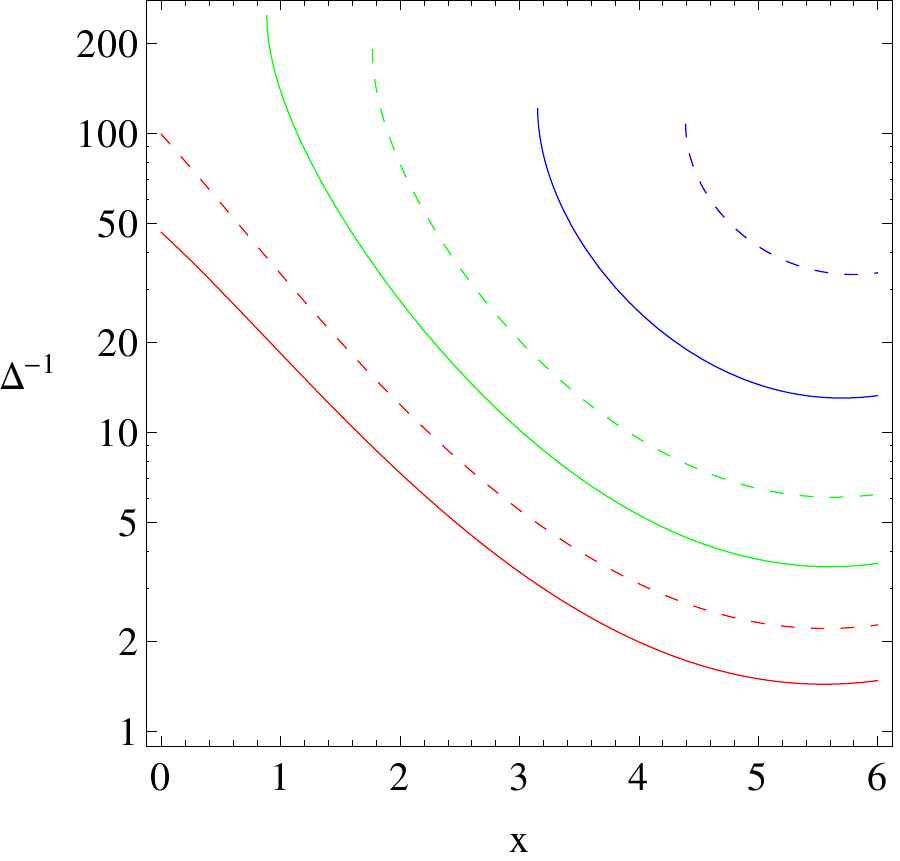}
}
\caption{$x$ axis: $x\equiv(A_t^2/\MS^2)$. The left panel shows the average stop mass $\MS$ required for constant Higgs mass $m_h$; the right panel shows the fine-tuning $\Delta^{-1}\equiv|\delta m_{H_u}^2| / (\tfrac{1}{2}M_Z^2)$ that results. The mediation scale $\Lambda$ is taken to be $10^5\,$GeV ($10^{16}\,$GeV would increase the fine-tuning by a factor $\sim\!6$.) Red curves (the lowest two) have $m_h=115$ GeV, green curves (the middle two) $m_h=119$ GeV, and blue curves (highest) $m_h=123$ GeV. Dashed (solid) lines have $\tan\beta=8$ ($30$). We take $M_t = 173.1$ GeV.}
\label{fig:ConstHiggsMasses}
\end{figure}

Secondly, one could invert the $\delta m_{h}^2$ expression to find the function $\MS(x)|_{\delta m_{h}^2}$ for how the stop mass varies as a function of $x$ for a constant Higgs mass.
This function is easily obtained from Eq. \eqref{CarenaWagner} which is a quadratic equation in $\log(\MS^2/M_t^2)$; we plot it in the left panel of Fig. \ref{fig:ConstHiggsMasses}.
The one-parameter function $\delta m_{H_u}^2(x,\,\MS(x)|_{\delta m_{h}^2})$ then gives the range of $\delta m_{H_u}^2$ possible for a given Higgs mass, depending on the amount of stop mixing ($x$) one uses to achieve that Higgs mass.
The {\it minimum} occurs at $x_{\text{natural}}$.
We plot this in the right panel of Fig. \ref{fig:ConstHiggsMasses}, normalised to $\tfrac{1}{2}M_Z^2$ for a transparent indication of fine-tuning.

The different colours (line styles) in Fig. \ref{fig:ConstHiggsMasses} correspond to different $m_h$ ($\tan\beta$), see the caption.
We see that the greater the $m_h$ we require (and the lower $\tan\beta$ is), the larger $x$ must be to even find a solution: no-mixing scenarios are more limited in the Higgs mass they can reach before the $m_h$ expression \eqref{CarenaWagner} breaks down.
Indeed even using the program FeynHiggs ~\cite{Heinemeyer:1998yj} for a higher-order calculation, in the no-mixing $x=0$ scenario breakdown occurs before one can reach $m_h\approx126$ GeV and one must resort to a matching of the MSSM on the SM, as noted in \cite{Draper:2011aa}.

The left panel of Fig. \ref{fig:ConstHiggsMasses} illustrates the obvious fact that the smallest stop mass for a given Higgs mass occurs at exactly maximal mixing $x=6$.
Close inspection of the right panel shows the more subtle point that the lowest fine-tuning occurs at {\it almost} maximal mixing.
We see from the flatness of the curve for $x\in[5,6]$, however, that the difference between the two is essentially nil.

\subsection{Finding {\it the} stop mass?}

\begin{figure}
\centering
\subfigure{
\includegraphics[width=0.47\linewidth]{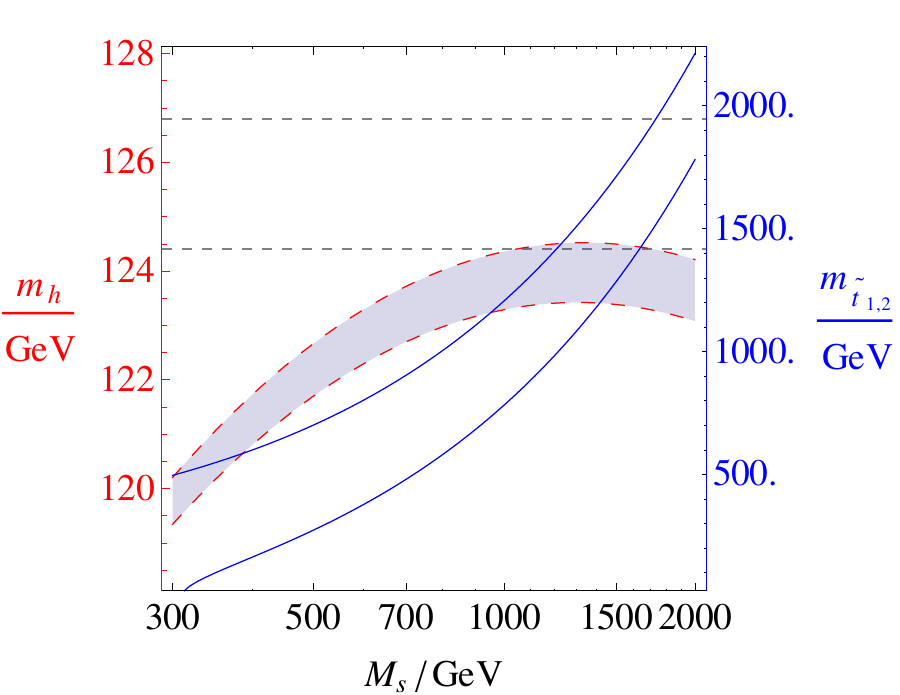}
}
\subfigure{
\includegraphics[width=0.47\linewidth]{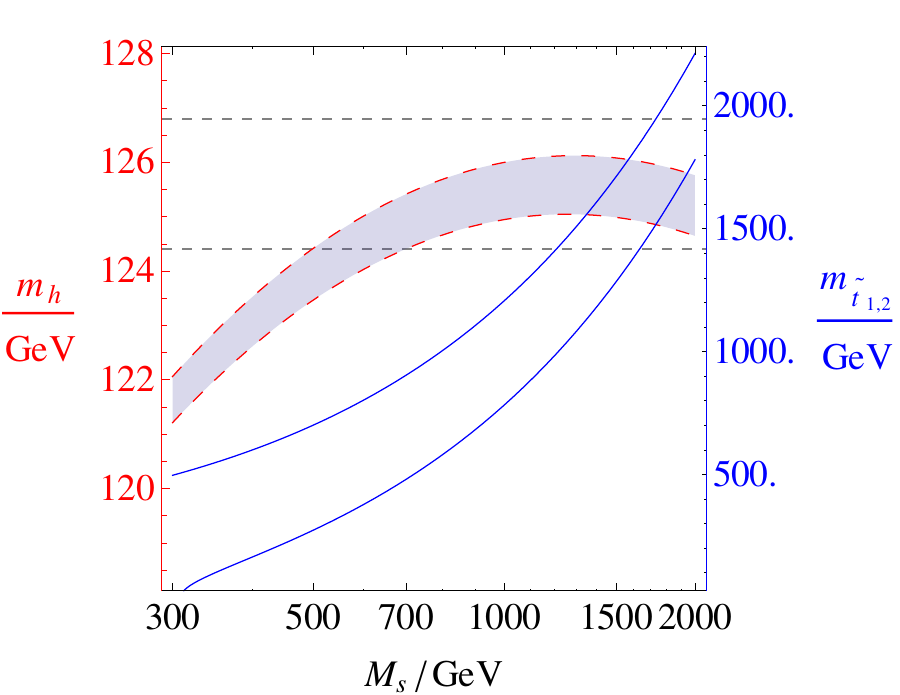}
}
\\
\subfigure{
\includegraphics[width=0.47\linewidth]{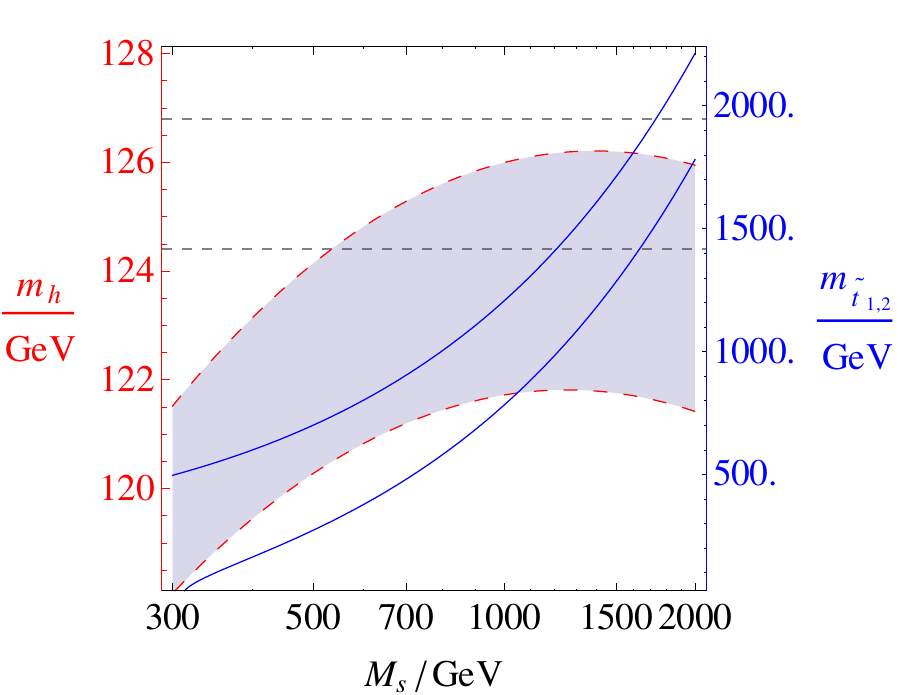}
}
\subfigure{
\includegraphics[width=0.47\linewidth]{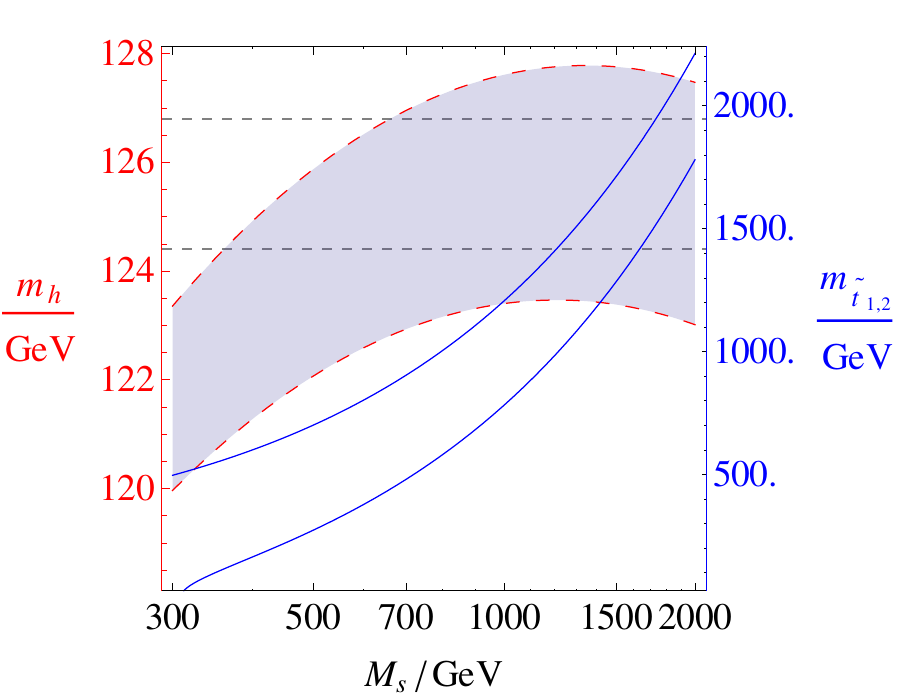}
}
\caption{Blue solid lines, right-hand $y$ axis: the tree-level stop mass eigenvalues $m_{\tilde{t}_{1,2}}$, assuming $m_{\tilde{Q}_3}^2 = m_{\tilde{u}_3}^2$.
Red dashed lines, left-hand $y$ axis: the two-loop expression of \cite{Carena:1995bx} for the mass of the lightest CP-even Higgs boson $m_{h}$, valid for $850\,\mbox{GeV} \lesssim \MS \lesssim 1500\,\mbox{GeV}$.
Further dashed lines indicate the lowest $m_h$ compatible with CMS's $m_h = (125.3 \pm 0.4_{\text{stat}} \pm 0.5_{\text{syst}})$ GeV and the highest $m_h$ compatible with ATLAS's $m_h = (126.0 \pm 0.4_{\text{stat}} \pm 0.4_{\text{syst}})$ GeV \cite{:2012gk,CMS}.
Curves are plotted as a function of the average stop mass $\MS$, with the ratio of stop mixing to stop mass taking its {\it most natural} value as defined in Eq. \eqref{MostNaturalRatio2loop} and plotted in Fig. \ref{fig:RatioNLO}.
Grey shading shows the Higgs mass uncertainty due to the top quark mass uncertainty.
Upper panels take the top quark pole mass as measured by the Tevatron, ATLAS and CMS: $M_t = (173.1 \pm 0.7) $ GeV  \cite{Degrassi:2012ry}; lower panels take $M_t = (173.3 \pm 2.8)$ as extracted from the Tevatron's $\sigma(pp\rightarrow t\bar{t}+X)$ measurement \cite{Alekhin:2012py}.
The left (right) panels are for $\tan\beta=8$ ($30$).}
\label{fig:HiggsMassesNLO}
\end{figure}

Varying $\MS$ while keeping $x=x_{\text{natural}}$ fixed traces out the Higgs mass that results in this most natural setting.
Of course to go to the full Higgs mass from only the stop radiative corrections one must either neglect the corrections from other sparticles (and so certainly steer clear of the large bottom-Yukawa regime at $\tan\beta \gtrsim \frac{m_t}{m_b}$), or else pick some `representative' value for all other sparticle masses and calculate their fixed contribution.
We do the former in Fig. \ref{fig:HiggsMassesNLO}. We will first explain the range of validity of Fig. \ref{fig:HiggsMassesNLO} before discussing the uncertainty arising from the top quark mass, shown with grey bands.

The Higgs mass expression \eqref{CarenaWagner} arises from the effective theory in which both stops have been integrated out at a single scale $\MS$, thus requiring $m_{\tilde{t}_1} \gtrsim \tfrac{3}{5}\,m_{\tilde{t}_2}$ \cite{Carena:1995bx} which results in a lower bound on $\MS$ for validity of the expression.
The lower bound is minimal when $m_{\tilde{Q}_3}^2 - m_{\tilde{u}_3}^2$ (which we have argued can be freely chosen) vanishes; we plot the resulting stop mass eigenvalues also in Fig. \ref{fig:HiggsMassesNLO}.
The bound $m_{\tilde{t}_1} \gtrsim \tfrac{3}{5}\,m_{\tilde{t}_2}$ can be seen to imply $\MS\gtrsim 850$ GeV.
\eqref{CarenaWagner} also does not contain higher-order terms $\mathcal{O}(\log^3(\MS^2/M_t^2))$, giving a corresponding upper bound for its validity.
Its accuracy is $\sim 2$ GeV for $\MS \lesssim 1.5$ TeV \cite{Carena:1995bx}.

Notice in Fig. \ref{fig:RatioNLO} that at $\MS\sim 1.3$ TeV, $x_{\text{natural}}$ becomes as high as $6$ (and takes higher values still for $\MS \gtrsim 1.3$ TeV).
This signals a breakdown in our procedure, since the Higgs mass expression is a symmetric function of $x$ about the value $6$, but naturalness always favours lower values to minimise $A_t$.
From Fig. \ref{fig:HiggsMassesNLO}, we see that at $\MS\sim 1.3$ TeV the derivative of the Higgs mass with respect to $\MS$ vanishes\footnote{
Note that the derivative of interest is $m_h$ with respect to $\MS$, with $x$ held constant; in Fig. \ref{fig:HiggsMassesNLO} the latter is {\it not} constant.
However it is varying sufficiently slowly that when we instead hold it exactly constant, the relevant derivative still vanishes at the same point $\MS\sim 1.3$ TeV.
}, which is purely an artefact of the truncated expression.
The Lagrange constrained optimisation, \eqref{LagrangeMin}, is then solved by the Higgs mass {\it alone} maximised with respect to both of its arguments, with the Lagrange multiplier $\lambda$ vanishing i.e. the naturalness consideration decouples.
Hence the solution is pushed onto exactly maximal mixing.
Even higher terms in the Higgs mass expression would be needed to push this breakdown point out to higher stop masses.

The authors of \cite{Degrassi:2012ry}, following a similar analysis to \cite{Bezrukov:2012sa}, take the top quark mass measurement relevant for calculation of the (SM) Higgs mass to be a combined measurement of the pole mass from the Tevatron, ATLAS and CMS: $M_t = (173.1 \pm 0.7) $ GeV.
In \cite{Alekhin:2012py} it was argued that direct experimental measurement of the top quark pole mass gives a theoretically ill-defined quantity, and that a more theoretically rigorous approach is to extract the running mass from measurement of the top pair production cross section, and thence obtain $M_t = (173.3 \pm 2.8)$ GeV.
We show both cases in Fig. \ref{fig:HiggsMassesNLO}; the choice of error in $M_t$ has a striking effect on the Higgs mass uncertainty.

An initial hope for this work was to see whether a given Higgs mass could give an indication of the average stop mass, using the principle of optimal naturalness to reduce the function\footnote{
Necessarily for a choice of $\tan\beta$; large but less than $\frac{m_t}{m_b}\sim 40$ gives us a maximal Higgs mass and without a fine-tuning penalty, which is clearly optimal.
}
\begin{equation} \label{1DHiggs}
m_h\; \approx \;m_h(\MS,x) \quad \rightarrow \quad\left.m_h(\MS)\right|_{x=x_{\text{nat}}}
\end{equation}
The latter is shown in Fig. \ref{fig:HiggsMassesNLO} with its uncertainty arising from the top quark mass uncertainty $\Delta M_t$; where it intersects with the observed Higgs mass, with its own error $\Delta m_h$, we see which stop masses are possible.
Fig. \ref{fig:HiggsMassesNLO} shows that even with the simplification of Eq. \eqref{1DHiggs}, the uncertainties $\Delta M_t$ and $\Delta m_h$ alone make any inference of $\MS$ from $m_h$ very difficult.
This is compounded by a theoretical uncertainty in the calculation of $m_h$, widely suggested to be 2-3 GeV, and the smaller contributions from the particles besides the stops.
The smallest $\MS$ compatible with the observed $m_h$, found for large $\tan\beta$ and conservative $\Delta M_t$, can be read off as $\sim 350$ GeV; however the top quark mass eigenvalues are then too split to trust a calculation based on integrating them both out at once (as discussed earlier).
Fig. \ref{fig:HiggsMassesNLO} also makes clear that even higher order terms than the two-loop corrections to $m_h$ are necessary to constrain $\MS$ from above, as the monotonic increase of $m_h$ with $\MS$ needs to be captured.
(An upper limit on $\MS$ {\it without} naturalness is given in \cite{Giudice:2011cg} -- a few $10^{8}$ GeV for split SUSY and unconstrained for high-scale SUSY; MSSM-to-SM matching is needed to calculate $m_h$ with stops far beyond the weak scale.)

We consider RG improvement to go beyond a leading log expression for $\delta m_{H_u}^2$ but relegate this discussion to the Appendix, as it is more involved though ultimately gives the same $x_{\text{natural}}$.

\section{Discussion and Conclusion} \label{Conclusion}

We have shown that {\it almost maximal mixing}, with $x\equiv A_t^2 / \MS^2$ slightly lower than $6$, is optimal; though we have also shown that the distinction between this case and maximal mixing $x=6$ is academic.
In other words to achieve a given Higgs mass $m_h$, balancing $A_t$ and $\MS$ to optimise naturalness gives almost the same result as simply trying to minimise $\MS$.

However conversely, even remaining in the MSSM, comparing how easily different models accommodate a $126$ GeV Higgs (a major focus of recent SUSY phenomenology) based on {\it how light the stops are} is misleading.
A maximal-mixing scenario will certainly have larger $m_h$ than a no-mixing scenario at the same $\MS$.
But note that the reasoning of the previous paragraph applies to a fixed mediation scale $\Lambda$.
If the maximal-mixing scenario has much larger $\Lambda$ than the no-mixing scenario -- e.g. if we take the former to represent supergravity (SUGRA) and the latter low-scale gauge mediation (GMSB) -- then it will be more unnatural not only due to the large $A_t$ but also due to large amount of RG running, i.e. the large logarithm in Eq. \eqref{deltamHuLL}.
Taking $\Lambda = 10^{16}\,$GeV and $10^{5}\,$GeV as representative of these two cases, the former will have an unnatural $\delta m_{H_u}^2$ term $\sim\!25$ times larger; the two should thus compare their Higgs masses with GMSB having stops (roughly $\surd 25$ times) heavier than SUGRA for similar fine-tuning, changing perhaps qualitatively the result of a comparison at fixed $\MS$, c.f. \cite{Arbey:2011ab,Arbey:2012dq}.
The optimal situation in the MSSM is clearly (nearly) maximal mixing with low-scale mediation: references \cite{Kang:2012ra} and \cite{Craig:2012xp} realised this with the introduction of large $A_t$ terms into gauge mediation via Higgs-messenger superpotential couplings. 

We hope that being analytic throughout, this work is complementary to the many numerical investigations of the Higgs in SUSY performed recently, showing more clearly the Higgs-stop-naturalness connection.
Two further sources of possible confusion in this area are as follows.
Maximal mixing is not {\it necessary} to achieve $m_h\approx126\,$GeV, as has been reported -- Fig. 6 of \cite{Draper:2011aa} for example shows that $\MS=\mathcal{O}(5\,\text{TeV})$ is sufficient with {\it no} mixing -- it is merely a less tuned method of doing so.
Also, we have shown that naturalness is insensitive to an arbitrary splitting of the stop masses $m_{\tilde{t}_2} - m_{\tilde{t}_1}$.
The corollary is that a light $\tilde{t}_1$ gives no indication of a natural theory, or an easy accommodation of $m_h\approx126\,$GeV.

Investigating whether minimal stop masses coincide with optimal naturalness, as done here for the MSSM, is particularly important for extensions of the MSSM.
Introducing a new particle which couples to strongly to the Higgs, in order to boost the latter's mass without heavy stops, is naively good for naturalness.
However pushing this new coupling as far as it will go can easily be imagined to introduce a new source of tuning somewhere in the theory (analogous to the effect of large $A_t$ on $\delta m_{H_u}^2$ considered here).
Exactly this effect in the Next-to-Minimal Supersymmetric Standard Model (NMSSM) was considered since this work first appeared, in \cite{Agashe:2012zq}: the stop masses and mixing were kept small, and the naturalness implications of NMSSM-specific contributions to $m_h$ calculated.
These were found to be a tuning of the lighter scalar's couplings to hide it from current collider constraints, in the case where the observed resonance is the second-lightest scalar; and worse tuning still when the resonance is the lightest scalar, to undo the `push-down' effect of level-repulsion between mass eigenvalues.

We have commented on the difficulty in tying down stop masses to a given Higgs mass, even invoking naturalness as a further constraining criterion.
The best chance then lies with specially designed searches for stops confirming their presence or absence at the weak scale.
We hope (against hope?) for the former.

\section*{Acknowledgements}
Thanks to the other participants of the ``Implications of a 125 GeV Higgs boson'' workshop held at LPSC Grenoble, where this idea was born, and to the Centre de Physique Th\'{e}orique de Grenoble for hospitality.
We thank Carlos Wagner and Thomas Rizzo for useful communications, and Matt Dolan, Joerg Jaeckel, Valya Khoze and Daniel Maitre for helpful comments.
Thanks also to Abdelhak Djouadi for encouragement on this work.
This work was supported by the STFC.

\appendix
\gdef\thesection{Appendix}
\section{Beyond Leading Log for $\delta m_{H_u}^2$}
The leading log expression for $\delta m_{H_u}^2$ is obtained by ignoring the scale dependence of $A_t$ and $\MS$; to do better we can consider integrating $A_t$ and $\MS$ over their varying higher-scale values.
First consider the running of $A_t$ and $\MS$ with arbitrary self and mutual couplings, as well couplings to other particles:
\begin{align} 
\frac{d}{dt}
\begin{pmatrix}
\MS^2(t) 
\\ A^2_t(t)
\end{pmatrix} = &
\begin{pmatrix}
a(t) & b(t) \\
0 & c(t)
\end{pmatrix}
\begin{pmatrix}
\MS^2(t)
\\ A_t^2(t)
\end{pmatrix}
+ \mbox{{\it other} running soft-mass terms} \label{RunningMixingStopsGen_Beta} \\
\therefore \quad 
\begin{pmatrix}
\MS^2(t) 
\\ A^2_t(t)
\end{pmatrix} = &
\begin{pmatrix}
d(t) & e(t) \\
0 & f(t)
\end{pmatrix}
\begin{pmatrix}
\MS^2(0)
\\ A_t^2(0)
\end{pmatrix}
+ \mbox{{\it other} high-scale soft-mass terms,} \label{RunningMixingStopsGen_Integrated}
\end{align}
where $a,b,c$ are running couplings and $d,e,f$ are related to the former by integration, and the lower-left entry of the matrix must vanish since $A_t$ appears in the Lagrangian, not $A_t^2$.
Note that if the {\it other} soft-mass parameters themselves run due to $A_t$ and $\MS$, this feeds back into Eq. \eqref{RunningMixingStopsGen_Integrated} as corrections to the coefficients $d,e,f$ at the next order in $g^2/16\pi^2$, which could have an impact but we neglect this for simplicity.
Integrating the $m_{H_u}^2$ beta function \eqref{mHuBetaFunction}, keeping just the stop-sector terms as before but now including their scale dependence (and that of the top Yukawa) as in \eqref{RunningMixingStopsGen_Integrated}, we have
\begin{multline} \label{Integrated_mHu}
\delta m_{H_u}^2(t) = \: \tfrac{3}{8\pi^2}\, \left( 2\MS^2(0) \int_{t'=0}^{t'=t} y_t^2(t') d(t')\:dt' + A_t^2(0) \int_{t'=0}^{t'=t} y_t^2(t') (2e(t')+f(t'))\,dt' \right) \\ + \mbox{{\it other} high-scale soft-mass terms}
\end{multline}
For Lagrange constrained optimisation, Eq. \eqref{LagrangeMin}, we must differentiate $\delta m_{H_u}^2$ and $\delta m_{h}^2$ with respect to $A_t^2$ and $\MS^2$.
One can either invert Eq. \eqref{RunningMixingStopsGen_Integrated} and substitute into Eq. \eqref{Integrated_mHu} to obtain $\delta m_{H_u}^2$ instead as a function of the {\it low}-scale stop parameters, or all derivatives can be taken with respect to the {\it high}-scale stop parameters (using the chain rule for $\delta m_{h}^2$, whose arguments should be evaluated at the low scale).
Both approaches give the same result, as they must:
\begin{multline}
f(t) \left( 2\int_{t'=0}^{t'=t}y_t^2(t') \, d(t') \:dt'\;  + d(t)y_t^2(t)\left( 1+\frac{A_t^2(t)}{2\MS^2(t)} \right) \right) \frac{\partial (\delta m_{h}^2)}{\partial A_t^2} = \\ \left( d(t)\int_{t'=0}^{t'=t}y_t^2(t') (2e(t')+f(t'))\,dt' 
 - 2e(t)\int_{t'=0}^{t'=t}y_t^2(t') \, d(t') \:dt' \right) \frac{\partial (\delta m_{h}^2)}{\partial \MS^2} \label{RGimprovedLagrangeMax_general}
\end{multline}
The leading log relation is recovered for $(d(t),e(t),f(t))=(1,0,1),\: y_t(t)=y_t$.
So what are these functions $d(t),e(t),f(t)$ in the MSSM?
Expressions for the one-loop running parameters can be written down when all Yukawa couplings except that of the top are set to zero \cite{Ibanez1984511,Essig:2007kh,Carena:1996km}.
$y_t$ and $A_t(t)$ do not require numerical integration if one also sets the $U(1)$ and $SU(2)$ gauge couplings to zero: one finds
\begin{gather}
y_t^2(t) = y_t^2(0) \,\xi^{-16/9}(t)\,G^{-1}(t;\tfrac{-16}{9}) \\
A_t(t)=G^{-1}(t;\tfrac{-16}{9})\left[A_t(0)+\frac{16}{9}M_3(0)\Big(G(t;\tfrac{-16}{9})\xi^{-1}(t)-\,G(t;\tfrac{-25}{9})\Big)\right] \label{runningA}  \\
\mbox{where}\quad\xi(t) = 1+\frac{3}{2\pi}\,\alpha_3(0)t \notag \\
G(t;n) = 1-\frac{3}{4\pi^2}\,y_t^2(0)\,\int_0^t\,dt'\,\xi^{n}(t') \notag 
\end{gather}
From \eqref{runningA} we read off that $f(t)=G^{-1}(t;\tfrac{-16}{9})$.
In this same scheme for extracting running parameters, the stop mass necessarily involves numerical integration.
However RG-induced splitting of the stop from the lighter-generation up-type quarks is typically small (and if not the model is necessarily unnatural, as mentioned in the Introduction), so that the running stop is well approximated by its high-scale value plus the gluino-induced term, the latter easily obtained by integrating the one-loop running gluino mass:
\begin{equation}
\MS^2(t) = \MS^2(0)+\frac{8}{9}M_3^2(0)\left( \frac{\alpha^2_3(t)}{\alpha^2_3(0)} -1 \right)
\end{equation}
This gives $d(t)=1,\:e(t)=0$.
Eq. \eqref{RGimprovedLagrangeMax_general} is then
\begin{multline}
2G^{-1}(t;\tfrac{-16}{9}) \left( \int_{t'=0}^{t'=t}\bigg( \xi^{-16/9}(t')\,G^{-1}(t';\tfrac{-16}{9}) \bigg) \,dt' \right. \\
\left. + \xi^{-16/9}(t)\,G^{-1}(t;\tfrac{-16}{9})\left( 1+\frac{A_t^2(t)}{2\MS^2(t)} \right) \vphantom{\int_{t'=0}^{t'=t}\bigg( \bigg)} \right) \frac{\partial (\delta m_{h}^2)}{\partial A_t^2} = \\
\int_{t'=0}^{t'=t}\bigg( \xi^{-16/9}(t')\,G^{-2}(t';\tfrac{-16}{9})\bigg) \,dt' \; \frac{\partial (\delta m_{h}^2)}{\partial \MS^2}
\label{RGimprovedLagrangeMax_MSSM}
\end{multline}
The integrals can be done analytically and the resulting root, $x_{\text{natural}}$, found; we do not plot it as it is essentially indistinguishable from the one shown in Fig. \ref{fig:HiggsMassesNLO}, even for very high mediation scales $\Lambda \sim 10^{16}$ GeV.
Thus our attempt at an approximate RG improvement of $\delta m_{H_u}^2$ (resumming all the logs that come with appreciable coupling constant factors) makes no difference to the result obtained from the leading log expression.

An alternative approach to this approximate RG improvement would be to work consistently at Next-to-Leading-Log NLL order for $\delta m_{H_u}^2$.
Barbieri-Giudice fine-tuning measures are given at NLL in \cite{CahillRowley:2012rv}, from which one can extract the dependence of $\delta m_{H_u}^2$ on any trilinear mixing term or sfermion mass-squared via
\begin{align}
 \delta m_{H_u}^2(A_i) = & \; \int \left( \frac{M_Z^2}{2A_i}  \left.Z_{A_i}\right|_{\tan\beta \rightarrow \infty} \right) \:d A_i \\
\delta m_{H_u}^2(m_{\tilde{f}}^2) = & \; \frac{M_Z^2}{4}  \left.Z_{m_{\tilde{f}}}\right|_{\tan\beta \rightarrow \infty} \label{eq:sfermionMassFT} \\
\mbox{where} \quad Z_{p_i} \equiv & \; \frac{\partial (\log M_Z^2)}{\partial (\log p_i)} \notag
\end{align}
Note that $\delta m_{H_u}^2$ and $m_{\tilde{f}}^2$ both having mass dimension $2$ results in $Z_{m_{\tilde{f}}} \propto m_{\tilde{f}}$, giving the simpler expression \eqref{eq:sfermionMassFT}. 
$Z_{A_i}$ however can contain further mass-scales beyond $A_i$; indeed for the stop, $Z_{A_t}$ contains terms with $M_1$, $M_3$ and $A_b$. In other words, at NLL $\delta m_{H_u}^2$ depends on $A_t$ not only through an $A_t^2$ term but also through terms $A_t M_1$, $A_t M_3$ and $A_t A_b$. In the spirit of connecting $\delta m_{H_u}^2$ and the Higgs mass to the stop sector in isolation, we will not explore this effect here.

\bibliographystyle{h-physrev3}
\bibliography{NaturalStops}

\end{document}